\documentclass[fleqn,usenatbib]{mnras}
\usepackage{newtxtext,newtxmath}
\usepackage{color,soul}
\usepackage[T1]{fontenc}
\usepackage{ae,aecompl}
\usepackage{graphicx}
\usepackage{amsmath,amssymb}
\title[Corrugated dust lanes]{Wobbly discs - corrugations seen in the dust lanes of edge-on galaxies}
\author[C. A. Narayan et al.]{
Chaitra A. Narayan,$^{1,2}$\thanks{E-mail: chaitra@ncra.tifr.res.in (CAN)}
Ralf-J\"urgen Dettmar,$^{2}$\thanks{E-mail: dettmar@astro.rub.de (RJD)}
Kanak Saha,$^{3}$
\\
$^{1}$National Centre for Radio Astrophysics, TIFR, Pune 411007, India  \\          
$^{2}$Ruhr-Universit\"at Bochum, Fakult\"at f\"ur Physik und Astronomie, Astronomisches Institut (AIRUB),
           Universit\"atsstrasse. 150, 44801 Bochum, Germany \\
$^{3}$Inter-University Centre for Astronomy and Astrophysics, Pune 411007, India
}

\date{Accepted XXX. Received YYY; in original form ZZZ}
\pubyear{2019}

\begin{document}
\label{firstpage}
\pagerange{\pageref{firstpage}--\pageref{lastpage}}
\maketitle           
           
\begin{abstract}

\noindent {We report the detection of small scale bending waves, also known as corrugations, in the dust lanes of five nearby edge-on disc galaxies. This phenomenon, where the disc mid-plane bends to become wavy, just as in warps but on a smaller scale, is seen here for the first time, in the dust lanes running across the discs. Because they are seen in absorption, this feature must be present in the dust disc in the outskirts of these galaxies. 
We enhance the visibility of these features using unsharp masking, trace the dust mid-plane across the disc, measure the corrugation amplitude by eye and the corrugation wavelength using Fourier analysis.  
The corrugation amplitude is found to be in the range of 70 - 200 pc and the wavelengths lie between 1 - 5 kpc.
In this limited sample, we find that the amplitude of the corrugations tends to be larger for lower mass galaxies, whereas the wavelength of corrugation does not seem to depend on the mass of host galaxies. Linear stability analysis is performed to find out the dynamical state of these dust discs. Based on WKB analysis, we find that the dust corrugations in about half of our sample are stable. Further analysis, on a larger sample would be useful to strengthen the above results.
}
\end{abstract}


\section {Introduction}
\label{sec:intro}

Many spiral galaxies, when viewed edge-on, are seen to host prominent dust lanes running across the mid-plane of their stellar discs \citep{Hackeetal1982, HowkSavage1999, Dalcantonetal2004, Holwerda+2019, mosenkov19}. Often, these dust lanes are thin in vertical extent, just obscuring the mid-plane of the host disc galaxy. This implies that the dust is kinematically cold and perhaps associated with the molecular gas including neutral hydrogen. In some face-on spirals, dust lanes are seen along the spiral arms, leading to the impression that dust is associated with lanes. Various observational analyses \citep{Nelson+1998, verstappen13, degeyter14, Smithetal2016} indicate however that the dust is diffusely distributed across a disc. So a dust lane seen across an edge-on disc is more likely a continuous disc rather than a ring-like structure . 
  
The formation of a well-defined dust lane itself is an exclusive feature of massive edge-on disc galaxies \citep{Dalcantonetal2004}. The presence of such a lane indicates that the ISM dust disc, dominated by large dust particles, experiences sufficient gravitational pull to settle down, is in vertical equilibrium and is no longer dominated by strong turbulence. If the turbulence in ISM were to dominate, the dust lane would be spread over a larger height about the galactic mid-plane. Even in the absence of strong turbulence, maintaining a long, thin dust lane across a disc is a challenging task. Bending instabilities could set in, either due to internal \citep[][and references therein]{Araki1985,ChequersWidrow2017} or external processes \citep[see][and references therein]{Weinberg1998,Gomez+2013,Widrow+2014}. So the state of dust lanes, whether they are 'thin and wavy' or 'thick and messy', is suggestive of what the disc is going through. 

 An axisymmetric disc when subject to a non-axisymmetric perturbation can produce bending waves about the mid-plane (z=0) such that at any given location in the disc ($R, \varphi$), the vertical displacement $\Delta z \propto \cos{m \varphi}$, m being the azimuthal wave number. The wave length of the bending waves should satisfy the $m \lambda = 2\pi R$ relation. $m=0$ refers to the bowl-shaped mode \citep{Sparke1995} while $m=1$ denotes the commonly observed warps. 
A number of physical mechanisms have been put forward to generate bending waves in galaxies, such as - tidal interaction with satellites or companion galaxies \citep{HunterToomre1969,Weinberg1998,WeinbergBlitz2006, Gomez+2013,Widrow+2014}; intergalactic matter accreted onto the dark halo \citep{JiangBinney1999} or directly onto the disc \citep{Lopez-Corredoiraetal2002}; intergalactic magnetic field \citep{Battaneretal1990} and intergalactic wind \citep{KahnWoltjer1959}. Bending waves are shown to also arise from various internal instabilities \citep{Araki1985,Revaz-Pfenniger2004,Sellwood1996,ChequersWidrow2017}, resonant coupling \citep{Binney1981}, dynamical friction between disc and halo \citep{NelsonTremaine1995} and halo substructure \citep{Chequersetal2018}.
In this light, the occurence of warps \citep{ReshComb1998,AnnPark2006} can be understood. The detection of higher order (large m and hence small amplitude) bending waves, however, remains largely under explored.

Bending waves with high azimuthal wavenumber ($m \sim 10$ or higher), also known as scalloping, were first observed in the very outskirts of our Galaxy \citep{Kulkarnietal1982,HJK1982} in the HI maps, and later in the CO \citep{Sandersetal1984}. More recently, \citep{Levineetal2006,KalberlaDedes2008} have confirmed the previous findings and also performed a detailed analysis of these $m=10 -15$ bending waves at around 30~kpc in the Milky Way. And they find that this feature is not a global one, i.e. the azimuthal bending waves are not seen across the entire disc at R=30kpc, but are rather limited to 
the direction of l=310$^{\circ}$. Throughout the rest of this paper, we refer to all small amplitude bending waves as corrugations. 

In this paper, we present the bf detection of corrugations in five nearby edge-on galaxies, as seen in their well-defined dust lanes in absorption against the background stellar disc light. The dust disc, with its scale height being smaller than the corrugation amplitudes, makes this detection possible.  We measure their wavelengths and amplitudes and present an analysis of the stability of these dust corrugations for our sample of galaxies.

The paper is organized in the following order.  In Section~\ref{sec:sample}, we describe our search for corrugation waves and sample selection. Section~\ref{sec:method} outlines the method employed in characterizing the corrugation waves.  Results and stability analysis for the waves are presented in Section~\ref{sec:results} and Section~\ref{sec:stability} respectively. Section~\ref{sec:discuss} describes the discussion over star formation connection and Section~\ref{sec:conc} presents the conclusions drawn from our analyses.

\section{Search for corrugations}
\label{sec:search}

\subsection{Why so few detected?}
\label{sec:sofew}

We note that so far, only four more galaxies apart from the Milky Way - NGC 4244, NGC 5907, NGC 5023 \citep{Florido+1991,Florido+1992} and IC 2233 \citep{MatthewsUson2008} show evidence for the presence of corrugations. 
This is surprising when we know that, (i) the m=1 warp has a high rate of occurence \citep{ReshComb1998,AnnPark2006}; (ii) the mechanisms that generate warps can also generate a host of other bending and breathing modes and (iii) it takes less energy to generate corrugations than warps.

The amplitude of corrugations is most likely the reason, for such low detection rates. 
Warps tend to show bending of a kpc and above. With the stellar disc thickness being less than a kpc (in late type galaxies where warps are mostly seen) this would be noticeable in nearby edge-on galaxies, despite flaring. 
Warps are more easily visible in the HI as HI generally extends farther out in a disc and the warp height increases with radius. 
On the other hand, the amplitude of corrugations is much smaller. 
For example, the radial bending wave in the inner Galaxy has a mid-plane deviation of only about 70pc and
wavelength of 2 kpc \citep{Quiroga1974} and 170pc at a radius of 14 kpc from the center \citep{Xu+2015}. 
Those seen in IC 2233, NGC 4244 and NGC 5023 are between 150pc - 200pc. Mid-plane undulations of this magnitude in stellar discs can be easily concealed due to a number of reasons.
First, the stellar disc scaleheight is larger than the corrugation amplitude. Second, while viewing an edge-on galaxy, our line-of-sight cuts across various radii in the disc. Thirdly, the differential precession ($\Omega -\nu/m$) of bending waves could erase them on rotation time scales (since m is large).

\subsection{A systematic search in the SDSS}
\label{sec:sdss}

As noticed before by \citet{MatthewsUson2008}, there has been no systematic search for corrugations in other galaxies, as yet.
Hence, our first step was to perform a systematic search for them using the SDSS archival data. Inspired by the dust lane in NGC 891 which shows tiny wiggles at the two ends, we selected a sample of nearby edge-ons (a/b > 5; a>1', Rmag <20, z=0) in the SDSS DR4. Out of the first 1000 edge-ons viewed by eye , barely 5-7 show apparent corrugations in the dust lane. This is < 1\% - much below expectation if the corrugations are believed to be generated by perturbation of a galactic disc via internal/external gravitational instability. Perhaps this rarity can be explained by the probability of a galaxy having all the following four factors - galaxy should be dust-rich; galaxy should show a continuous dust lane running along the mid-plane; galaxy is mildly perturbed; the generated corrugations are not damped yet.

To increase the probability of spotting corrugations, we then looked at a classified catalog of edge-ons that are more likely to be under the influence of gravitational instability. \cite{Luttickeetal2000} provide a complete sample of 1350 edge-on disc galaxies derived from the RC3 Catalogue that include a 45\% of boxy-peanut bulges.  This ensures a near 50\% presence of gravitational instability like a bar in the sample.  We made a subset of 358 galaxies from this list that have SDSS data (in DR4).  Of these, 91 show well-defined dust lanes and 68 show corrugations in them. Thus if we consider the well-defined dust lane galaxies as the parent sample, occurrence of corrugations in this sample is as high as 75\%. This is very similar to the occurrence of warps in late type edge-on galaxies \citep{ReshComb1998, GarciaRuizetal2002, AnnPark2006}

\subsection{Sample selection}
\label{sec:sample}
\cite{Luttickeetal2000} have compiled a complete sample of $\sim 1350$ near edge-on (60$^{\circ}$ $\leq$ i $\leq$ 90$^{\circ}$) galaxies from the RC3 catalogue of bright galaxies \citep{devauc1991} with D$_{25}$ > 2', in order to classify bulges.
We made a subset of this list consisting of 358 galaxies which have SDSS data.
182 of the above are selected that are highly inclined (i $>$ 80$^{\circ}$) plus dust rich.
About 68 of these galaxies show vivid corrugations in their dust lanes. A lot of these though are rather local, i.e. limited to a small segment of the dust lane.
For this pilot study, we have chosen a sample of five edge-on dust rich galaxies that show a clear wavy pattern of dust distribution along most of the galaxy mid-plane (see Table~\ref{tab:paratab1}). 
Of these, four are from the SDSS database and one (IC 2531) from the ESO NTT located at La Silla and observed with the instrument EMMI. 
Details relating to the obsevation of IC 2531 can be found in \cite{RossaDettmar2000,Pohlenetal2000}. 
IC 2531 is also one of the seven edge-on galaxies observed for their dust lanes by the HERschel Observations of Edge-on Spirals (HEROES) project by \cite{verstappen13}. 
We note that NGC 4013 of their sample also shows partial corrugation in its dust lane, barely visible in the SDSS but more clearly seen in a recent HST image.

\begin{figure*}
\rotatebox{0}{\includegraphics[height=7.0cm]{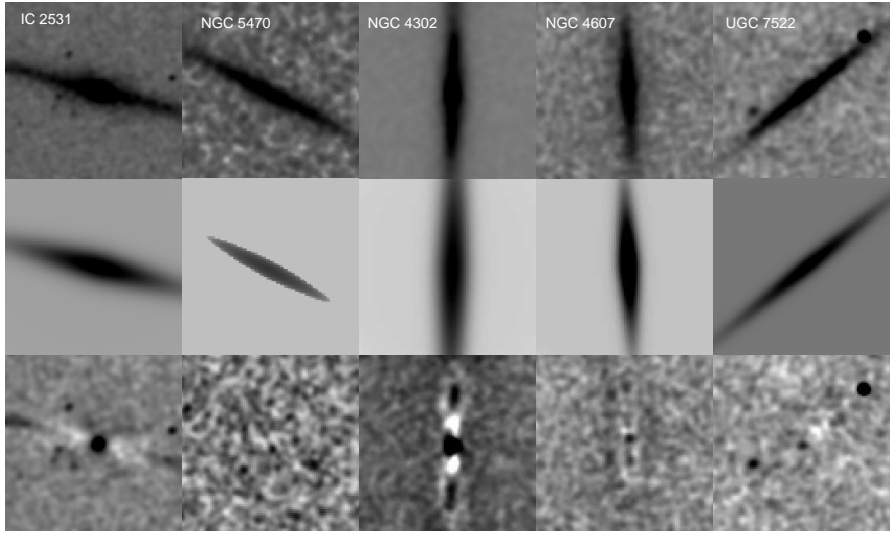}}
\caption{GALFIT performed on the 2MASS Ks band images of the galaxies in our sample. The upper panel refers to the 2MASS images, the middle row shows the GALFIT models and the lower ones are the residuals. }
\label{fig:2mass}
\end{figure*}

\section {Method}
\label{sec:method}
For each of the five galaxies in our sample, we trace the corrugations seen in their dust lanes and characterize them by measuring their amplitude and wavelength as discussed below. We also determine the stellar disc parameters of these galaxies using GALFIT on the 2MASS Ks band images. 

\begin{figure*}
\rotatebox{0}{\includegraphics[height=5.0cm]{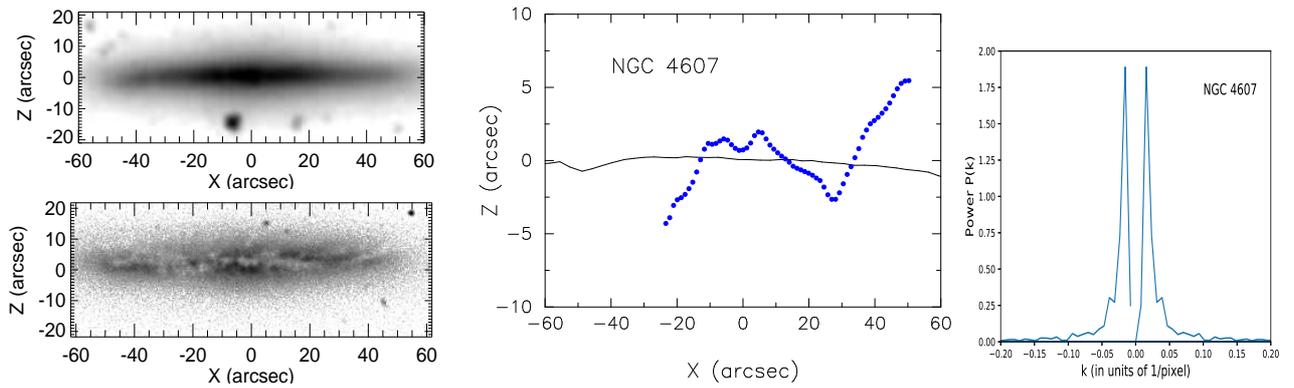}}
\caption{{\bf NGC 4607}: The top left panel shows the Spitzer/IRAC 4.5$\mu$ image; bottom left is the SDSS R-band image after unsharp masking. 
The mid panel shows the location of the stellar mid-plane (solid line) and the dust mid-plane (blue filled circles) derived from the left panel images as explained in Section~\ref{sec:measurecorr}. 
The right panel shows the result of Fourier analysis on the dust mid-plane. 
The peak on the right panel corresponds to corrugation wavelength of $\sim$ 3.1 kpc. 
1 arcsec $\sim$ 77 pc.}
\label{fig:ngc4607}
\end{figure*}

\begin{figure*}
\rotatebox{0}{\includegraphics[height=5.0cm]{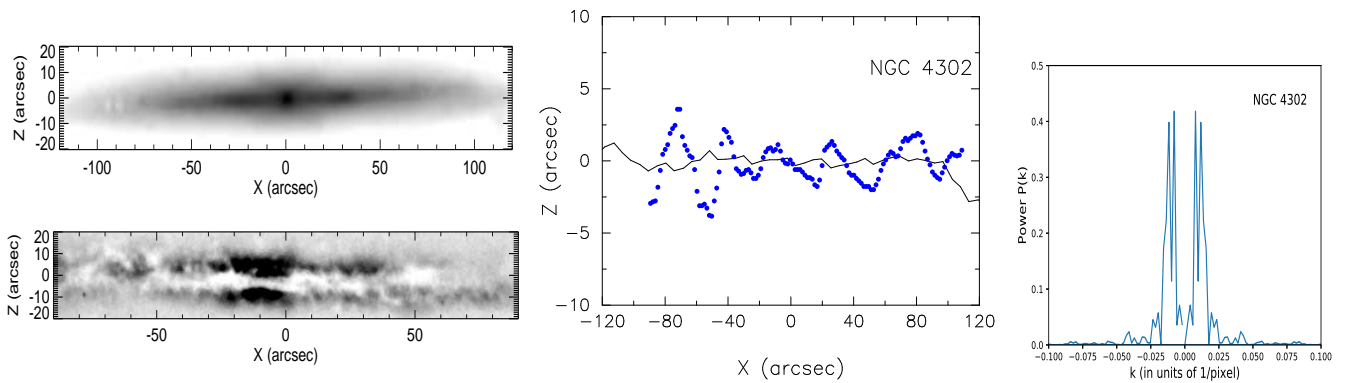}}
\caption{{\bf NGC 4302}: Details on the figure panels are as described in the caption of Fig \ref{fig:ngc4607}. 
The mean wavelength corresponding to the two nearby peaks on the right panel is $\sim$ 3 kpc. 1 arcsec $\sim$ 76 pc. We note that NGC 4302 is undergoing tidal interaction with NGC 4298, which might have caused the perturbations as found in \citet{SchwarzkopfDettmar2001}.}
\label{fig:ngc4302}
\end{figure*}


\begin{figure*}
\rotatebox{0}{\includegraphics[height=5.0cm]{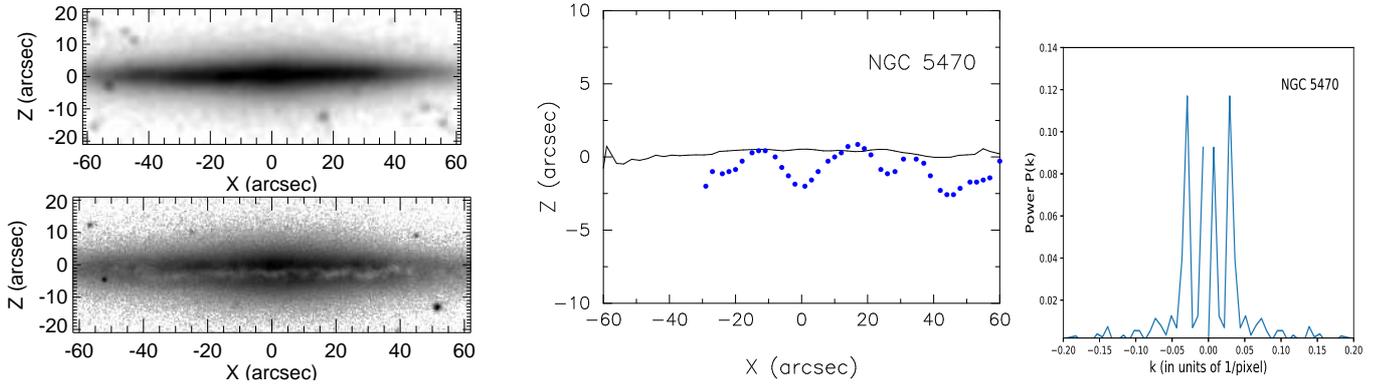}}
\caption{{\bf NGC 5470}: Details on the figure panels are as described in the caption of Fig \ref{fig:ngc4607}. 
The taller peak on the right panel corresponds to $\sim$ 1.6 kpc. 1 arcsec $\sim$ 119 pc. 
The dust lane is clearly offset from the stellar mid-plane perhaps due to the galaxy not being a perfect edge-on.}
\label{fig:ngc5470}
\end{figure*}

\begin{figure*}
\rotatebox{0}{\includegraphics[height=5.0cm]{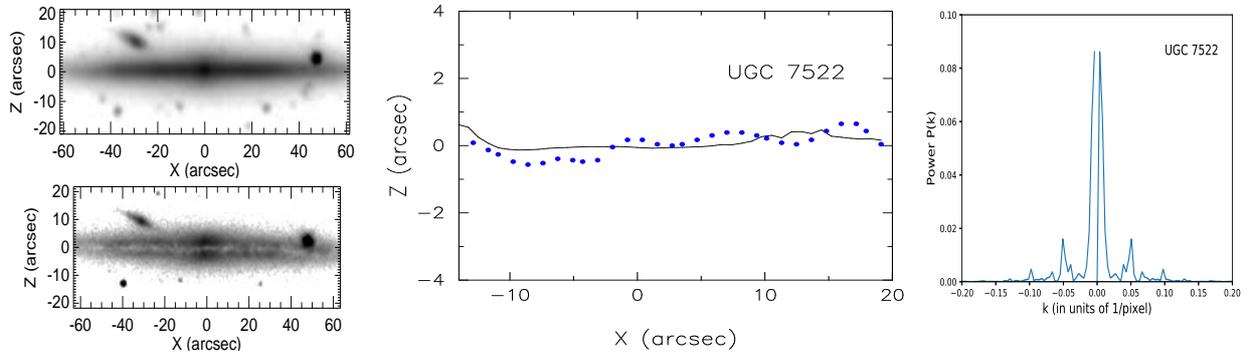}}
\caption{{\bf UGC 7522}: Details on the figure panels are as described in the caption of Fig \ref{fig:ngc4607}. 
The first peak on the right panel corresponds to $\sim$ 13 kpc. We consider the corrugation wavelength to be $~\sim$ 1 kpc which corresponds to the second peak. 1 arcsec $\sim$ 117 pc.}
\label{fig:ugc7522}
\end{figure*}

\begin{figure*}
\rotatebox{0}{\includegraphics[height=5.0cm]{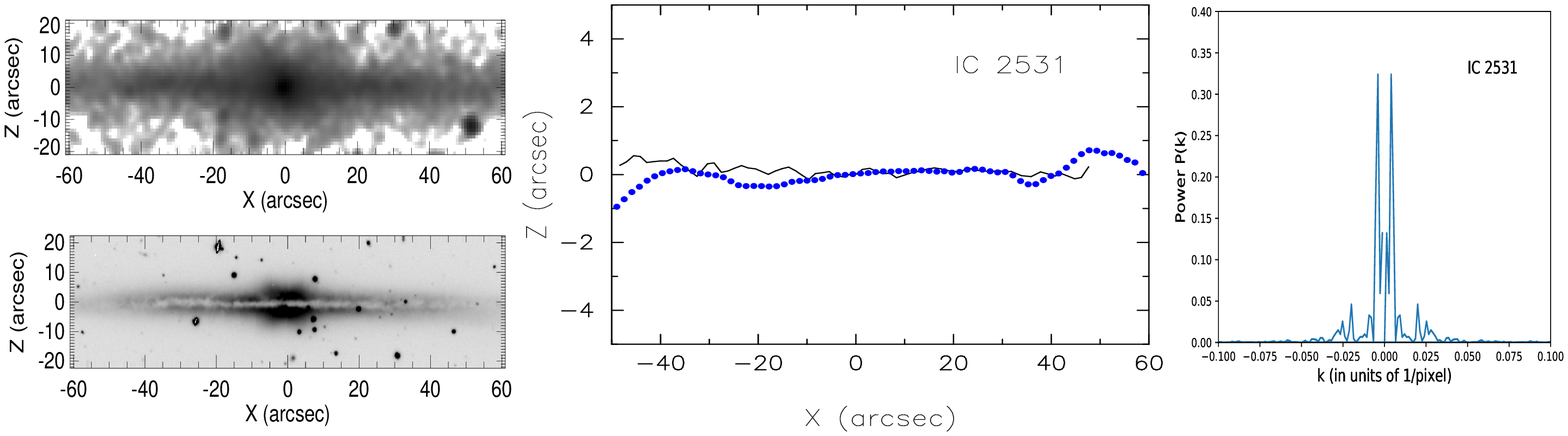}}
\caption{{\bf IC 2531}: The top left panel shows the 2MASS image; bottom left is the ESO-EMMI R-band image after unsharp masking. 
The most prominent peak on the right panel corresponds to a wavelength of $\sim$ 5 kpc. 1 arcsec $\sim$ 159 pc.
This is a massive galaxy with a boxy/peanut bulge. Note that the dust lane is aligned with the stellar mid-plane. }
\label{fig:ic2531}
\end{figure*}

\subsection {Measuring corrugations}
\label{sec:measurecorr}

We use the technique of 'unsharp-masking' that is known to enhance the visibility of fine structures in images. Each image is then rotated so that the mid-plane is horizontal and each column of pixels corresponds to a vertical cut across the galactic disc. The light distribution in each column typically shows two peaks separated by a central trough. We identify the trough as due to the foreground dust absorbing the background starlight and its central minimum as the centre of the vertical dust distribution. The identification of central minimum is done by eye using the pixel table feature in DS9. By tracing these minima across the entire disc, ie, along the length of the dust lane, we trace the mid-plane of the dust disc, seen in extinction (shown as blue dots in the mid-panel of Figures 2-6). The solid black lines are the stellar disc mid-planes derived from the 2MASS/Spitzer/IRAC data for each galaxy. For this, the vertical distribution of stellar disc is fit with a single Gaussian and the position of the peak intensity is traced across the disc by repeating the Gaussian fitting on all columns \citep{Sahaetal2009}. It should be noted that the black line is coming from the total intensity along each line-of-sight across the disc mid-plane, whereas the blue dots are indicative of the shape of the mid-plane at the disc edges facing the observer. Deviation from exact edge-on inclination of a galaxy might result in an offset between the two mid-planes. 

The outer discs of all the five galaxies display clear corrugations. UGC 7522 and IC 2531 show sub-arcsecond mid-plane deviations but NGC 4607, NGC 4302 and NGC 5470 show greater waviness. We use Fourier transform to extract the wavelengths of these patterns (see the right panel of Figs 2-6). Sometimes, more than one peak (the observed corrugation could be a superposition of two or more wavelengths) is seen. We then take the wavelength corresponding to the strongest peak as our best measurement. The amplitude of corrugation is measured by eye - it is taken to be half of the difference between peak to trough of the largest wave. Table 2 gives the wavelength and amplitude derived for all the sample galaxies.

\subsection{Surface photometry}
\label{sec:surfphot}

We perform photometric analysis to derive the structural parameters for our sample galaxies using GALFIT \citep{Pengetal2002}. 
The stellar surface brightness is modelled with two components - bulge and disc, for which we fit a sersic profile and an exponential profile respectively. 
The mathematical expression for these profiles are given below.

\noindent For the bulge we have,

\begin{equation}
I_b(r,n) = I_{b0} e^{[-k((r/R_e)^{1/n} -1)]}
\label{eq:bulgelum}
\end{equation}

\noindent where $I_{b0}$ is the central surface brightness of the bulge; $R_e$ and $n$ denote the 
effective radius and sersic index. The parameter $k$ depends on $n$. For further details, see
\cite{Pengetal2002}.

\noindent For the stellar disc,

\begin{equation}
I_d(r,z)= I_{s0}(r/R_s) K_{1}(r/R_s) sech^2(z/z_s)
\label{eq:disclum}
\end{equation}

\noindent where $I_{s0}$ is central surface brightness of the disc, $R_s$ and $z_s$ are the disc scale length and scale height. 
The function $K_1$ refers to the modified Bessel function.

The primary aim of performing surface photometry is to get the parameters e.g., scale length ($R_s$), scale height ($z_s$) for the stellar discs in our sample galaxies. 
We choose the 2MASS Ks band image for doing so as they are readily available for all the galaxies in our sample and at 2.2$\mu$, the dust extinction is lower. This would give a fair estimate of the disc parameters. 
We present our GALFIT analysis in Fig.~\ref{fig:2mass}. 
While fitting we have let all the parameters of the bulge and disc free. 
GALFIT does fit the stellar discs pretty well except for large residuals in the central regions of NGC 4302 and IC 2531. These two galaxies host a bar and a boxy-peanut bulge and hence a simple two-component model will not suffice.
Since our primary concern here is to derive a reasonable estimate of disc parameters, we refrain from improving bulge fits. 
We note that for NGC 4302, the $R_s$ derived using the above method is very close to that obtained from a double exponential profile fit to its Spitzer/IRAC image \citep{Sahaetal2009}.

\section {Results}
\label{sec:results}

\begin{table}
\centering
\caption{Parameters of the stellar discs derived by GALFIT for our sample are listed in the top panel. Below that
we tabulate the same for four more galaxies where corrugations have been previously quantified. We note
that seven out of nine galaxies (except IC 2531 and the Milky Way) are part of the $S^4G$ sample 
(see \citet{Butaetal2015}). The distances are taken from HYPERLEDA. References: Milky Way - \citet{Meraetal1998}; 
IC 2233 - \citet{MatthewsUson2008} NGC 4244 \& NGC 5023 - \citet{Sethetal2005}}
\begin{tabular}{ccccc}  \hline\hline 
Galaxy name  & $D$    & $R_s$  & $z_{s}$ & Features \\
       & (Mpc) & (pc) & (pc)  \\
\hline
\hline                     \\

NGC 4607  &17.3 & 1208.9 & 300.3  & interacting \\
NGC 4302  &17.0 & 2387.0 & 454.3  & bar + BP bulge \\
NGC 5470  &24.6 & 2616.0 & 456.0  & no bulge/bar \\
UGC 7522  &24.4 & 2655.9 & 409.5  & no bulge/bar \\
IC 2531   &31.0 & 3664.0 & 1072.0 & bar + BP bulge \\
\\
\hline                          \\

Milky Way & 0.0  & 3200.0 & 300.0  & bar \\
IC 2233	  & 10.0 & 1400.0 & 240.0  & bar + BP bulge\\
NGC4244   & 5.0  & 1780.0 & 257.0  & no bar \\
NGC5023   & 8.0  & 1280.0 & 160.0  & no bar \\ 

\hline
\end{tabular}
\label{tab:paratab1}
\end{table}

\begin{table}
\centering
\caption[ ]{Corrugation parameters. The last two columns of the top panel of the table list the corrugation parameters as derived in Sec \ref{sec:measurecorr}. We also list other important host galaxy properties like the maximum rotation velocity and the stellar disc mass. $V_{max}$ is taken from HYPERLEDA.
In the lower panel, the same parameters are compiled from the literature for the four other galaxies.
References: NGC 5023 \& NGC 4244 - \cite{Florido+1991}; IC 2233 - \cite{MatthewsUson2008}; Milky Way - \cite{Quiroga1974}.}

\begin{tabular}{cccccccc}  \hline\hline 
Galaxy name      & $V_{max}$ & $M_{star}$ & $\lambda_{corr}$ & $\Delta{z}_{corr}$ \\
        & (kms$^{-1}$) &$(\times 10^{10} M_{\odot})$ & (pc) & (pc) \\
\hline
\hline                     \\

NGC 4607   & 98.9  & 0.427 & 3103.1 & 200.2 \\
NGC 5470   & 110.1 & 0.466 & 1570.0 & 300.0 \\
UGC 7522   & 139.3 & 1.681 & 1029.6 & 70.2   \\
NGC 4302   & 167.6 & 3.524 & 3025.0 & 292.6 \\
IC 2531    & 228.0 & 12.07 & 4992.0 & 115.2  \\
\\
\hline                                             \\

NGC5023    & 80.3  & 0.185 & 1150.0 & 140.0\\ 
IC 2233	   & 85.0  & 0.233 & 7000.0 & 150.0 \\
NGC4244    & 98.0  & 0.412 & 2863.0 &  198.0\\
Milky Way  & 220.0 & 6.43 & 2000.0 & 70.0 \\

\hline
\end{tabular}
\label{tab:paratab2}
\end{table}

Table~\ref{tab:paratab1} lists the stellar disc parameters derived from GALFIT as mentioned in Sec~\ref{sec:surfphot}.
The measured wavelengths and amplitudes of corrugations (see Sec~\ref{sec:measurecorr}) are shown in Table~\ref{tab:paratab2}.
Table~\ref{tab:paratab1} and Table~\ref{tab:paratab2} list the parameters not only of our sample of five galaxies, but also of galaxies with previously detected corrugations. This serves as a single-point reference for the rare sightings of corrugations and allows for easy comparison of host galaxy properties. However, we note a few inhomogeneities here. For example in Table~\ref{tab:paratab2}, the disc component showing corrugation varies from dust to HI to stellar disc. The method used to derive the corrugations parameters may also differ. Yet, we hope that this naive comparison will lead to more homogeneous data sets in future.

Our analyses on each of the five sample galaxies are shown through Fig.~\ref{fig:ngc4607} to Fig.~\ref{fig:ic2531}. 
Corrugations are prominent in NGC 4302, NGC 5470 and UGC 7522. 
On the other hand, in IC 2531, the dust lane is pretty much straight and aligned with the stellar mid-plane in the central parts, becoming progressively wavy outwards. 
In the case of NGC 4607, we do not have a well defined dust lane running all along the galaxy mid-plane, as a result of which only a smaller segment of corrugation is seen.

\subsection{Dependence of corrugation wave length on rotation velocity}
\label{sec:rotvel}

\begin{figure}
\rotatebox{270}{\includegraphics[height=9.0cm]{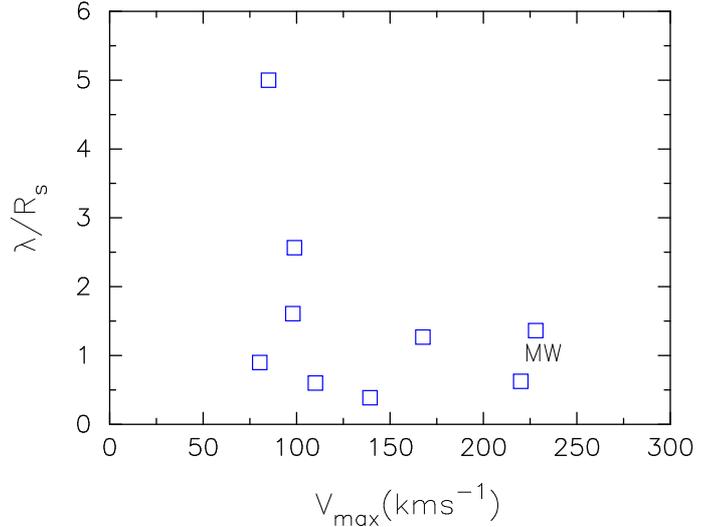}}
\caption{Dependence of corrugation wavelength normalized by the radial scale length of the stellar disc on the peak rotation velocity of host galaxy. }
\label{fig:wavelength}
\end{figure}

The measured corrugation wavelengths for the entire sample range from about $1 - 7$~kpc. 
It is interesting to note that the typical epicyclic amplitude e.g., in our Milky Way is $\sim 1$ kpc. 
Here we study the dependence of the size of these waves on the peak rotation velocity of the host galaxy. This is to check whether the host galaxy has any preference in the size of the bending waves that are allowed to stay.
In Fig.~\ref{fig:wavelength}, we show the variation of the wavelengths normalized by their respective disc scale lengths, with the peak rotation velocity. 
The galaxies in our sample span a fairly wide range of rotation velocities ranging from about $80$~kms$^{-1}$ to $230$~kms$^{-1}$. 
We see that, for heavier galaxies the corrugation wavelengths are nearly equal to the disc scale length. And for low mass galaxies, there is no such preference.  
We are yet to understand the generality of this result which requires performing the same analysis on a larger sample of galaxies.

\subsection{Dependence of corrugation amplitude on the stellar disc mass}
\label{sec:stellarmass}

\begin{figure}
\rotatebox{270}{\includegraphics[height=9.0cm]{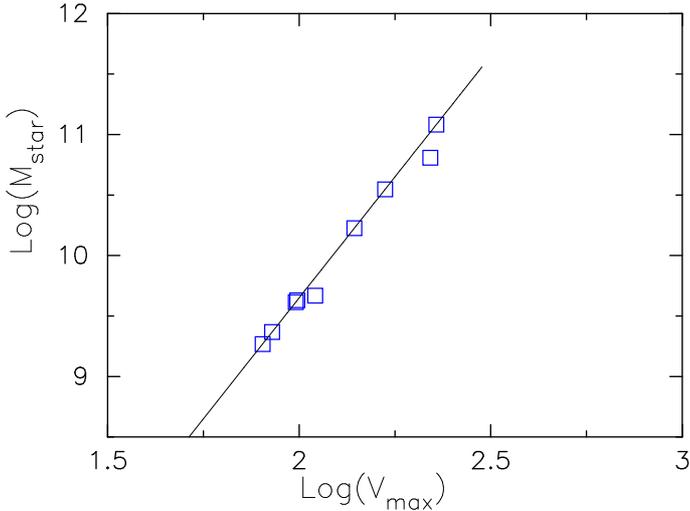}}
\caption{Stellar mass Tully-Fisher relation for our sample of  galaxies. The solid line represents the model that we have used to determine the stellar masses. }
\label{fig:TF}
\end{figure}

\begin{figure}
\rotatebox{270}{\includegraphics[height=9.0cm]{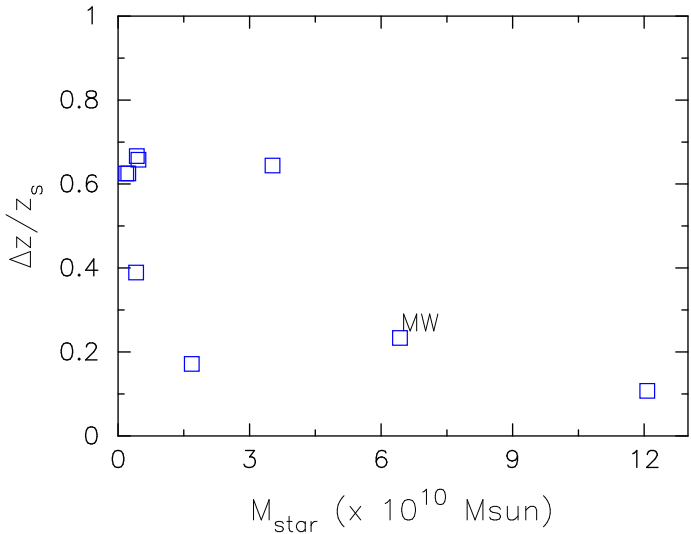}}
\caption{Dependence of the corrugation amplitude normalized by the stellar scale height on stellar mass of the host galaxy.}
\label{fig:amplitude}
\end{figure} 

\cite{Dalcantonetal2004} find that prominent thin dust lanes are often seen in massive edge-on disc galaxies. 
In low-mass galaxies, dust distribution is not confined to a lane but is spread to larger heights above and below the mid-plane. 
Following this trend, perturbed low-mass galaxies could be hosting taller corrugations than perturbed massive galaxies. 

In order to determine the total stellar masses for our sample galaxies, we employ the stellar mass Tully-Fisher (TF) relation  \citep{Bell-deJong2001, Courteauetal2003, McGaugh2005, Kassinetal2007}. 
In particular, we use the relation $M_{star}=44.66 V_c^4$ \citep{McGaugh2005} to determine the total stellar masses 
(see Table~\ref{tab:paratab2}). 
This has also been depicted in Fig.~\ref{fig:TF}.
Note that our approach to derive an estimate of the total stellar mass may not be accurate enough but this serves the present purpose, as we are primarily interested in getting a general idea about the dynamical nature of these corrugations.

As the stellar mass of a disc galaxy increases, the gravitational force on the disc components towards the mid-plane increases. This effect is perhaps the reason for the appearance of well-defined dust lanes in heavier edge-on galaxies and a messy (unsettled) dust distribution in lower mass edge-ons \citep{Dalcantonetal2004}. It is worth mentioning here that decades ago, \cite{Hackeetal1982} had found that the thickness of dust lane correlated with the galaxy luminosity. Hence it is of great interest to see if the amplitude of the corrugations seen here depends on the galaxy mass as well.

In Fig.~\ref{fig:amplitude}, we plot the amplitude of the corrugations vs the stellar mass of the host galaxy. 
The amplitude is normalized by the scale height of the stars. 
As expected, larger amplitude waves are, indeed seen in low-mass galaxies and the corrugation amplitude decreases as the stellar mass increases.
Again, a future analysis on a larger sample could confirm this trend.
It is interesting to note that the corrugation amplitude is typically confined well within the scale height of the stars and this confinement is stronger in massive stellar discs. 
In addition, we also compute a quantity called $\zeta_{nlin}\equiv \Delta{z}_{corr}/\lambda_{corr}$ for our sample galaxies. 
When the values of $\zeta_{nlin} \ge 0.5$, non-linearity might become important to understand these bending waves. 
But for all the sample galaxies, we have $\zeta_{nlin} < 0.1$, ensuring that a linear stability analysis would be appropriate for now.

\section{Stability of dust disc}
\label{sec:stability}
In order to perform stability analysis, we need to understand the dust distribution, its total mass and its velocity dispersion in our sample galaxies. 
Most of these are local galaxies within about $20 -30$~Mpc.
We assume that the dust properties, especially the dust density distribution does not vary much, within these galaxies. 

\subsection{Dust distribution and velocity dispersion}
\label{sec:veldisp}

The truly 'hamburger' like edge-ons i.e., one with an apparently continuous dust lane running across the disc diameter is certainly indicative of a dust rich galaxy.
The dust seen over the galaxy centre could actually be the dust at the physical edge of the disc that happens to be in the line of sight towards the centre.
We assume that these long thin dust lanes are actually edge-on projections of a thin dust disc. 
We further assume for simplicity that such a dust disc is rotating with the gas and stars and has its density distributed like that of stars i.e. an exponential radial profile with either $sech^2$ or Gaussian like vertical profile \citep[]{Bianchi2007, Smithetal2016, casasola17}, although more recently \citep{mosenkov19} have found a significant deviation from an exponential radial distribution. 

Unlike neutral hydrogen gas, the velocity dispersion ($\sigma_{dust}$) of dust is complicated because of the varying grain size and their relative coupling to the interstellar medium (ISM). 
Larger grains are less coupled to the gas and move basically under the gravitational forces; they are the ones confined extremely close to the mid-plane. 
For these larger grains, the velocity dispersion is typically $\sim 1 - 2$~kms$^{-1}$. 
Whereas lighter grains are coupled to the motion of the gas and attain a velocity dispersion which is closer to the turbulent velocity of the ISM, as a result of which they are spread to larger heights above the mid-plane ($z=0$).
The gas velocity dispersion has been observed to be nearly constant with radius at approximately $9 - 10$~kms$^{-1}$ in our own Galaxy \citep{Spitzer1978, Malhotra1995}.
In a large number of external galaxies, the gas velocity dispersion is shown to be in the range of $7 - 9$~kms$^{-1}$ \citep{Lewis1984, Kamphuis1993} consistent with that of our Galaxy.
We consider an intermediate value of $5$~kms$^{-1}$ which is more like an average between the heavy and lighter grains; this is also the velocity dispersion of molecular hydrogen gas in our Galaxy \citep{Clemens1985}.
Although it is not very relevant here to distinguish dust based on their grain sizes, it is extremely important to keep that in mind while drawing any conclusion on their stability.

\subsection{Dispersion relation for corrugations}
\label{sec:disprel}

Corrugations are seen in stars, gas and dust indicating that they are most likely caused by gravitational instabilities.
\cite{SchwarzkopfDettmar2001} find that most interacting galaxies show vertical perturbations and flaring in their stellar discs. 
Via simulations, \cite{WidrowBonner2015} show that a satellite galaxy or a dark matter sub-halo that passes through a stellar disc may excite vertical oscillations in the disc.

Since the inner region of a galactic disc is resistant to strong perturbation as shown theoretically \citep{SahaJog2006,pranav2010} and observationally in the HI disc of our own Galaxy by several authors \citep{Quiroga1974,Kulkarnietal1982,Spicker1986,Levineetal2006},  we consider these corrugations as arising in the outer parts of a disc.
But unlike the HI gas, the dust disc is in general radially confined within the extent of the stellar disc. 
So the typical size of a dust disc could be $\le R_{25}$ where $R_{25}$ is the radius at which the stellar disc surface brightness drops to $25$~mag/arcsec$^2$. 
In terms of disc scale length, $R_{25} \sim 4 -5 R_s$. 
From \citep{HunterToomre1969}, the azimuthal wavenumber $m$ of corrugations in the dust disc could be:

\begin{equation}
m \simeq {2 \pi} \frac{R_{25}}{\lambda_{corr}}
\label{eq:waveno}
\end{equation}

In the above relation, if we choose $R_{25}=4 R_s$ and $\lambda_{corr} \sim 1.5 R_s$, the azimuthal wavenumber at the edge of stellar/dust disc could be as high as $m \sim 16$. Such deductions are consistent with high-m bending waves found in our Galaxy \citep{Levineetal2006}. Clearly, $m$ depends on the size of the dust disc and wavelength of corrugation. 

Now the question is whether these corrugations could go unstable?
We consider a simple WKB bending wave model here. Then, the vertical displacement of the mid-plane of the dust disc is:

\begin{equation}
\Delta{z}(r,\varphi) = \Re{H(r)e^{ikr}\times e^{i(\omega t - m\varphi)}}
\label{eq:bending}
\end{equation} 

\noindent where $k$ denotes the wave number and $\omega$ is the frequency. 

Following \cite{HunterToomre1969}, the dispersion relation for the corrugation is:

\begin{equation}
{\omega_d}^2 = \nu^2(r) + 4\pi^2 G \frac{\Sigma_{dust}(r)}{\lambda_{corr}} - 4\pi^2 \frac{\sigma_{dust}^2}{\lambda_{corr}^2}
\label{eq:disp_full}
\end{equation}

\noindent where ${\omega_d} = \omega - m \Omega$; $\Omega$ and $\nu$ are the local azimuthal and vertical frequency of the galactic disc; $\Sigma_{dust}$ is the surface density of dust and $\sigma_{dust}$ is the dust velocity dispersion.

\noindent Since the corrugations are likely to be in the outskirts where the stellar disc self-gravity itself becomes less important, we can safely ignore the self-gravity of the dust disc. 
In that case, the local vertical frequency which has its dominant contribution from the surrounding dark matter halo, can be approximated as $\nu = V_c/(q r)$, where $V_c$ is the rotation velocity and $q$ is the halo flattening. 
Then Eq.~\ref{eq:disp_full} can be written as:

\begin{equation}
{\omega_d}^2 = \frac{V_c^2}{q^2 r^2} - 4\pi^2 \frac{\sigma_{dust}^2}{\lambda_{corr}^2}
\label{eq:disp_simp}
\end{equation}

The above relation can be used to directly compute the critical wavelength for corrugations by the setting $\omega_d^2 = 0$. 
In doing that, we replace $V_c = V_{max}$ and $r=5 R_s$, then for a given flattening of the dark halo, we have,

\begin{equation}
\lambda_{crit} = 10 \pi q \frac{\sigma_{dust}}{V_{max}/R_s}
\label{eq:lamdacrit}
\end{equation}

Any wavelength larger than the critical wavelength $\lambda_{crit}$ would be stable in the galaxy. 
Although there is no general proof of stability for bending waves of arbitrary wave length, it can be shown that low m bending waves (namely m=1 warp or m=0 bowl shape mode) are stable in cold self-gravitating discs \citep{HunterToomre1969,Saha2008} and these are the largest wavelength bending waves. 
From our simplified analysis, it is interesting to note that any corrugation whose wavelength is greater than the scale length of the underlying stellar disc, is found to be stable.

\begin{table}
\centering
\caption[ ]{Stability parameters for the corrugations listed in Table \ref{tab:paratab2}}
\begin{tabular}{ccccc}  \hline\hline 
Galaxy name  & $\lambda_{corr}/R_s$ & $\omega_d^2/\Omega_d^2$  & $\lambda_{crit}/R_s$ \\
       & &  &  \\
\hline
\hline                     \\

NGC 4607  &2.57 & 0.0341 & 1.429  \\
NGC 4302  &1.26 & 0.0274 & 0.843  \\
NGC 5470  &0.60  &-0.176 & 1.284 \\
UGC 7522  &0.38 & -0.289 & 1.015  \\
IC 2531   &1.36 & 0.0391 & 0.620  \\
\\
\hline                          \\

Milky Way & 0.62 & -0.0028 &0.642  \\
IC 2233	  & 5.0  & 0.0439 & 1.663 \\
NGC4244   & 1.61 & 0.0096 & 1.442 \\
NGC5023   & 0.90 & -0.1402 & 1.760\\ 

\hline
\end{tabular}
\label{tab:paratab3}
\end{table}

\begin{figure}
\rotatebox{270}{\includegraphics[height=9.5cm]{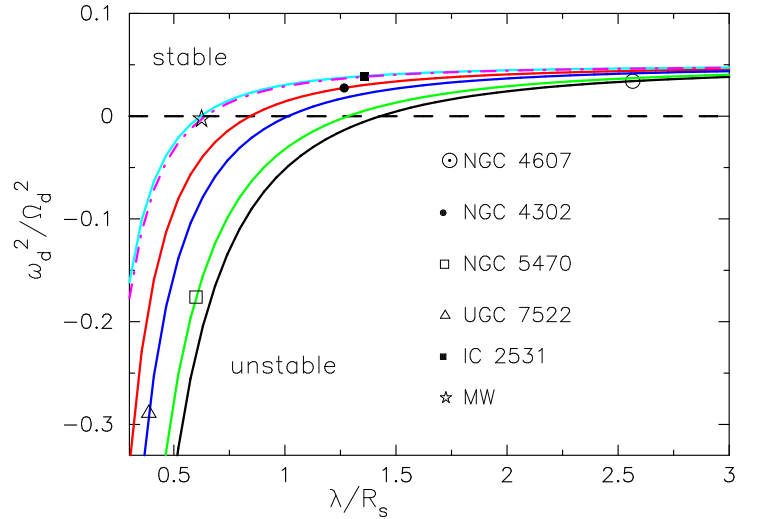}}
\caption{Stability diagram for the dust corrugations in our sample galaxies. 
The horizontal dashed line represents the marginal stability curve for these waves. Black - NGC 4607, red - NGC 4302, green - NGC 5470, blue - UGC 7522, cyan - IC 2531 and pink dash-dot - Milky Way. 
Note, $\Omega_{d}= V_{max}/R_s$ }
\label{fig:disp}
\end{figure}

\section {Discussion}
\label{sec:discuss}
\subsection{Star formation connection}
\label{sec:starform}

Corrugations seen in stars and gas  can lead to very different consequences. 
While the former can dissipate by heating up the stars, the latter can lead to star formation, thus making the phenomenon itself hold a key role in the secular evolution of disc galaxies. 
Corrugations have been linked to the formation of dense gas condensations and therefore with regulation of 
star formation \citep{NelsonMatsuda1980,Alfaroetal1992}.
Initial theoretical attempt to explain corrugations was made by \cite{Nelson1976a,Nelson1976b} using linear perturbation analysis. 
In a follow-up non-linear analysis by means of simulations, \cite{NelsonMatsuda1980} showed that the vertical oscillations in the gas disc of a spiral galaxy leads to clumping of gas. 
\cite{Klessenetal2000} show an upper limit for the wavelength of perturbation that can lead to cloud collapse. 
These findings suggest that the corrugation phenomenon is not just a minor detail in the structure of gas layers but is dynamically very significant as it can lead to star formation in the outskirts of the disc. 

H$\alpha$ imaging of NGC 891 by \cite{Dettmar1990} shows clear corrugations in the outer edges of the disc suggesting connection between star formation and corrugations.
\cite{Alfaroetal1992} find the density of young open clusters as well as GMCs correlate with vertical displacement from the galactic plane along the Carina-Sagittarius arm. 
Examining the morphology of H$\alpha$ regions, they suggest a relation between the corrugations and star formation. 
They suggest that the phenomena responsible for corrugation do not necessarily trigger star formation, instead they redistribute the interstellar medium creating conditions favourable for it. 
More recently \cite{Elmegreenetal2014} find star formation happening in the dust lanes of S4G galaxies. 
Some face-on disc galaxies like M94, NGC 278, NGC 1512 and NGC 4314 show outer rings of star formation. It would be very interesting to check if corrugations are at work here.

\subsection{Corrugations in the Milky Way}
\label{sec:MW}
Our position in the Galaxy allows us to deduce the corrugations in the Milky Way in great detail.
They have been seen in the inner Galaxy ever since the first HI maps were made \citep{Gumetal1960, Quiroga1974}. The same was observed in HII regions
and supernova remnants \citep{Lockman1977} and in CO \citep{CohenThaddeus1977}. Similar 'waviness' of the mid-plane was noticed along the spiral arms in B and OB stars \citep{Dixon1967,lynga1970}; and HI \citep{Spicker1986}.

With the advent of stellar kinematics surveys like RAVE, GAIA, SDSS, LAMOST etc, we are able to not only map the Galactic stellar disc in great detail but also peek into our Galaxy's past (using the velocity information).
Structures related to bending and breathing modes have been identified in our Galactic stellar disc from within the solar radius \citep{Carrillo+2019} upto the edges of the disc. 
\cite{Xu+2015} identify four substructures using SDSS DR8: two peaks of corrugations (bending modes) at 10.5 kpc and 14 kpc; and two rings (breathing modes) - the Monoceros ring at 16.5 kpc and the TriAnd ring at about 21 kpc from the galactic center. Gaia DR2 has led to the discovery of more such substructures in the Galaxy - e.g., vertical bending modes \citep{Bennett&Bovy2019}, phase-space spirals \citep{Antoja+2018} that are a manifestation of bending waves \citep{Darling+2019}, asymmetries in stellar velocity distribution \citep{Carrillo+2019} etc.
Bending waves exist beyond the stellar disc as well. At about the radius of 30 kpc, \cite{Levineetal2006} and \cite{KalberlaDedes2008} find azimuthal  corrugations (with high azimuthal wave number, m=10-13) in the HI disc.

Simulations have endorsed the picture that most of these spatial and velocity substructures are due to bending and breathing modes that might have been caused from the impact of Sagittarius dwarf falling into our Galactic disc \citep{Gomez+2013,Carrillo+2019}. Further \cite{Gomez+2017} have shown that bending and breathing modes must be a common feature in most disc galaxies.

\subsection{De-projection of corrugations seen in edge-on galaxies}
\label{sec:deproj}

Simulations \citep{EdelsohnElmegreen1997,Gomez+2013} show that the bending waves generated by a Sagittarius-like perturber has both radial and azimuthal dependence.
So far corrugations in other galaxies have all been detected in edge-on discs (all galaxies in Table \ref{tab:paratab1} except Milky Way) where we see the combined effect of radial and azimuthal dependence. Looking for corrugations in less inclined discs is not an option, unless we look in the velocity space \citep{Alfaroetal2001,Sanchezetal2015}.
However \cite{Kamphuis+2013} combine the HALOGAS data with non-axisymmetric modeling (using a version of TIRIFIC) and deproject HI observations of edge-on galaxy NGC 5023. They  find that their data is best explained by an edge-on disc showing radial corrugations (see their Fig 10). Hence we stress the importance of non-axisymmetric modeling here as the need of the hour because, with the availability of high-resolution HI data on scores of nearby edge-on galaxies, a statistical study of corrugations is only a small step away.

\section {Conclusion}
\label{sec:conc}
The present study has focused mainly on characterizing corrugations in the dust lanes of a sample of five edge-on galaxies and understanding their physical state. 
Partial dust waves are seen in many galaxies but only a handful of galaxies show waves long enough to be characterized. 
This paper is one of the first attempts to give a quantitative estimate of their size and scale as well as address the issue of whether they are stable. 
In order to strengthen our analyses, we have added four more galaxies including the Milky Way where corrugations have been previously measured in stellar and gaseous discs. 
The fact that corrugations are seen in stars, gas and dust  leads to the possibility of gravitational instability as the likely origin. 
 In fact, 5 of the 9 galaxies that we have tabulated are visibly perturbed (bar/tidal interaction). 

\vspace{0.5cm}
\noindent Our main conclusions from this work are the following:

1. We present the first systematic study of dust corrugations in edge-on galaxies and characterize them by measuring their wavelengths and amplitudes. 

2. The amplitude of corrugation seems to depend on the host galaxy mass.

3. The wavelength of corrugation does not show any dependence on host galaxy mass. It is roughly about the stellar disc scalelength for most of our sample.

4. By carrying out a stability analysis, we show that the corrugations with wavelengths larger than the scalelength of the stellar disc, are stable. 

5. We find indications that corrugations can be as common as warps. We also find mounting evidence that corrugations play a key role in the growth of disc galaxies by aiding star formation in its outskirts.

\section*{Acknowledgements}
\noindent We thank the referee for carefully reading the manuscript and providing suggestions that has greatly improved the quality of the paper. Most of the above work was done while CAN was a Humboldt Fellow at the AIRUB, Bochum. 
CAN would like to thank the Alexander von Humboldt Foundation for the fellowship and AIRUB for hosting the fellow. 
The authors would like to thank  G. Aronica, J. van Eymeren, V. Heesen, K. Polsterer, O. Schmithuesen, T. Falkenbach, D. Bomans, Y. Shekinov, P. Kalberla, C.J. Jog, R.R. Deshpande and S. Sridhar for useful discussions.

Funding for the SDSS and SDSS-II has been provided by the Alfred P. Sloan Foundation, the Participating Institutions, the National Science Foundation, the U.S. Department of Energy, the National Aeronautics and Space Administration, the Japanese Monbukagakusho, the Max Planck Society, and the Higher Education Funding Council for England. 
The SDSS Web Site is http://www.sdss.org/.

The SDSS is managed by the Astrophysical Research Consortium for the Participating Institutions. 
The Participating Institutions are the American Museum of Natural History, Astrophysical Institute Potsdam, University of Basel, University of Cambridge, Case Western Reserve University, University of Chicago, Drexel University, Fermilab, the Institute for Advanced Study, the Japan Participation Group, Johns Hopkins University, the Joint Institute for Nuclear Astrophysics, the Kavli Institute for Particle Astrophysics and Cosmology, the Korean Scientist Group, the Chinese Academy of Sciences (LAMOST), Los Alamos National Laboratory, the Max-Planck-Institute for Astronomy (MPIA), the Max-Planck-Institute for Astrophysics (MPA), New Mexico State University, Ohio State University, University of Pittsburgh, University of Portsmouth, Princeton University, the United States Naval Observatory, and the University of Washington.

This research has made use of the NASA/IPAC Extragalactic Database (NED), which is operated by the Jet Propulsion Laboratory, California Institute of Technology, under contract with the National Aeronautics and Space Administration.
This research has made use of NASAs Astrophysics Data System bibliographic services. 
This research has made use of SAOImage DS9, developed by Smithsonian Astrophysical Observatory.

\label{lastpage}
\end{document}